\documentclass[showpacs,preprintnumbers,superscriptaddress,amsmath,reprint,aps,prl]{revtex4-1}
\usepackage{bm}
\usepackage{graphicx} 
\usepackage{color}
\usepackage{amsmath}
\usepackage{mathtools}
\usepackage{booktabs}
\usepackage[thinc]{esdiff}
\usepackage{threeparttable}
\usepackage{flafter}
\usepackage{xcolor}

\begin{document}

\title{On the universality of spectroscopic constants of diatomic molecules}

\author{Xiangyue Liu}

\affiliation{Fritz-Haber-Institut der Max-Planck-Gesellschaft, Faradayweg 4-6, 14195 Berlin, Germany }

\author{Gerard Meijer}

\affiliation{Fritz-Haber-Institut der Max-Planck-Gesellschaft, Faradayweg 4-6, 14195 Berlin, Germany }

\author{Jes\'{u}s P\'{e}rez-R\'{i}os}

\affiliation{Fritz-Haber-Institut der Max-Planck-Gesellschaft, Faradayweg 4-6, 14195 Berlin, Germany }

\date{\today}

\begin{abstract}

We show, through a machine learning approach, that the equilibrium distance, harmonic vibrational frequency, and binding energy of diatomic molecules are universally related. In particular, the relationships between spectroscopic constants are valid independently of the molecular bond. However, they depend strongly on the group and period of the constituent atoms. As a result, we show that by employing the group and period of atoms within a molecule, the spectroscopic constants are predicted with an accuracy of $\lesssim 5\%$. Finally, the same universal relationships are satisfied when spectroscopic constants from {\it ab initio} and density functional theory (DFT) electronic structure methods are employed.

\end{abstract}

\date{\today}

\maketitle

Early in the history of molecular spectroscopy, when it became a discipline within chemical physics in the 1920’s ~\cite{Herzbergbio}, some intriguing empirical relationships between different spectroscopic properties were observed~\cite{Kratzer1920,Mecke1925,Morse1929}. In particular, it was found that the equilibrium distance, $R_e$, and the harmonic vibration frequency, $\omega_e$, were correlated in diatomic molecules. As the field evolved, the relationship between $R_e$ and $\omega_e$ became more evident, and more empirical relations between spectroscopic constants were identified~\cite{Clark1934bis,Badger1934,Clark1935,Clarck1934,Gordy1946,Guggenheimer1946,Clark1941,Clark1941bis}. However, these empirical relationships were typically only valid for certain atomic numbers or groups of the constituent atoms. This result motivated the development of realistic diatomic molecular potentials~\cite{Morse1929,Linnett1940,Newing1940,Linnett1942,Varshni1957,Varshni1958} and triggered the physical chemistry community to think about the ``periodicity'' of diatomic molecules~\cite{Hefferlin}.

The development of quantum chemistry assisted in the understanding of the physics behind the empirical relationships between spectroscopic constants. In particular, thanks to the application of the Hellmann-Feynman theorem, it was possible to connect $\omega_e$ directly with the electronic density at $R_e$~\cite{Salem1963,Kim1964,Borkman1968,King1968}. As a result, a first principles-based explanation of the observed empirical relations between spectroscopic constants appeared~\cite{Borkman1969,Politzer1970,Anderson1969,Anderson1970,Anderson1971,Simons1971,Anderson1972,gazquez1979universal}. Nevertheless, the obtained relations based on the electronic density were only valid for subsets of molecules, and to date it has not been possible to find universal relations between atomic identifiers and for spectroscopic constants of diatomic molecules.


\begin{figure}
    \centering
    \includegraphics[width=1\linewidth]{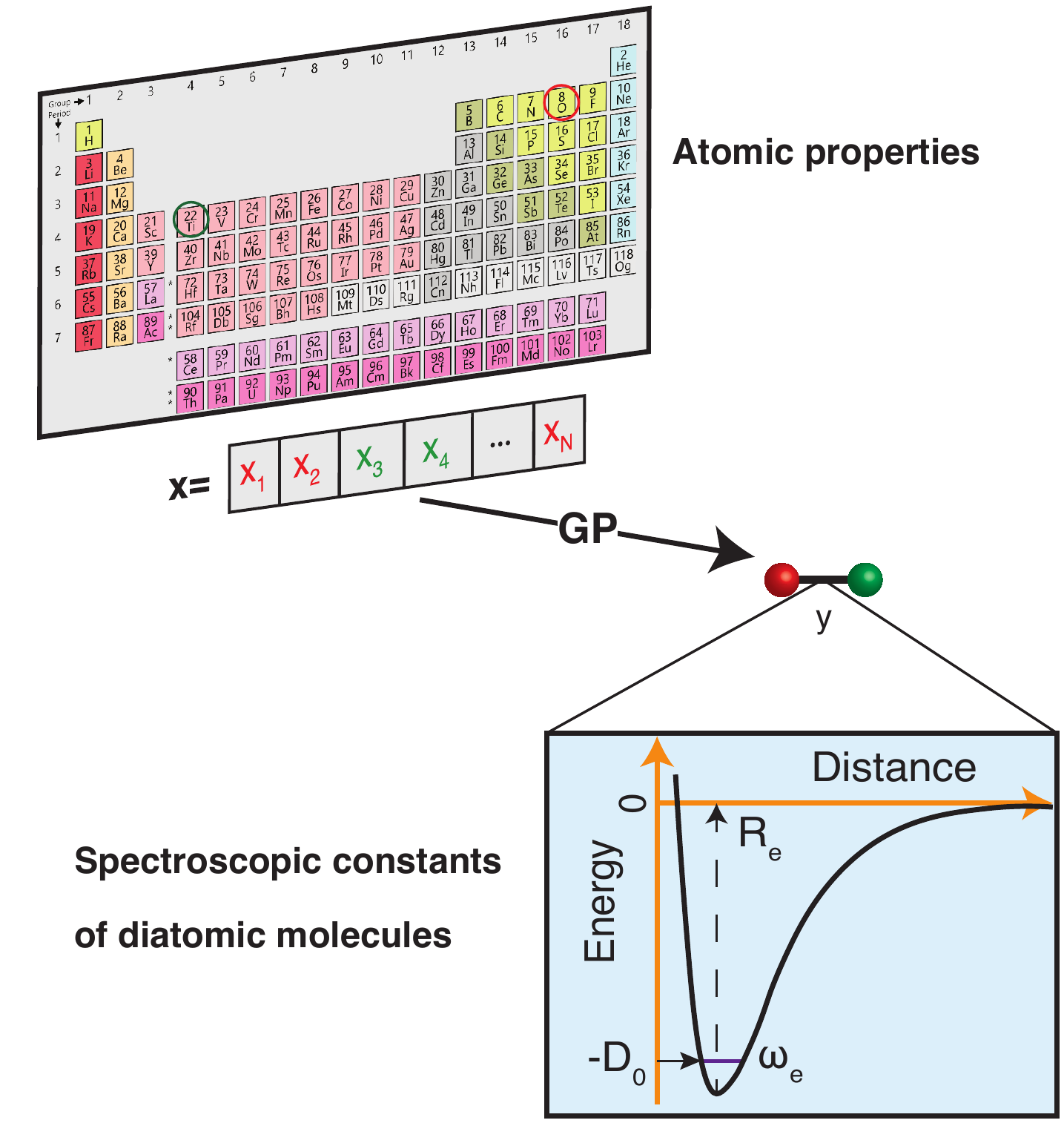}  
    \caption{ Universal relationships between spectroscopic constants. Sketch of the Gaussian process (GP) regression model for spectroscopic constants of diatomic molecules as a function of atomic properties of the atoms within a molecule.}
    \label{Fig:Concept}
\end{figure}

In this letter, we show that the relationship between spectroscopic constants of diatomic molecules is independent of the kind of molecules at hand and hence universal. Our findings rely upon the application of state-of-the-art non-linear machine-learning (ML) models to an orthodox dataset of spectroscopic constants for diatomic molecules. In particular, we apply the Gaussian process (GP) regression model~\cite{williams2006gaussian} to predict $R_e$, $\omega_e$, and the binding energy, $D_0$, as a function of the groups and periods of the constituent atoms. As a result, it is possible to predict those spectroscopic constants with an accuracy of $\lesssim 5\%$. Finally, we show that the spectroscopic constants coming from high-level electronic structure methods (density functional theory (DFT) and {\it ab initio}) display the same relationships as the experimental data employed.


The quest for universal relationships between spectroscopic constants is related to the problem of how atomic and molecular properties describe a spectroscopic property of a molecule, $y=f(\mathbf{x})$. Here, $\mathbf{x}=(x_1,x_2,...,x_n)$, consists of different atomic properties of the constituent atoms or molecular properties, as presented in Fig.~\ref{Fig:Concept}, where $n$ denotes the number of input features relevant for the problem at hand. Unlike traditional (non-)linear regression models, which assume a fixed form of function $f(\mathbf{x})$, GP embraces a Bayesian perspective and presumes a prior distribution over the space of functions $f(\mathbf{x}_i) \sim \mathcal{GP}(m(\mathbf{x}_i), K(\mathbf{x}_i, \mathbf{x}_j))$ with a joint multivariate-Gaussian distribution, centered at $m(\mathbf{x_i})$ and characterized by the covariance function $K(\mathbf{x}_i,\mathbf{x}_j)$, which specifies the correlation (or ``similarity'') between data points~\cite{williams2006gaussian}.

In this work, the spectroscopic properties $y$ are modeled as

\begin{equation}
 \label{Eq:GP_py}
    P(y_i | f(\mathbf{x}_i), \mathbf{x}_i) \sim \mathcal{N} (y_i| \mathbf{h}(\mathbf{x}_i)^T {\beta} + f(\mathbf{x}_i), \sigma_y^2).
\end{equation} 

\noindent
where the basis functions, $\mathbf{h}(\mathbf{x}_i)$, project $\{\mathbf{x}_i\}$ to a new (higher dimensional) feature space with coefficients $\beta$, and $\sigma_y$ includes the noise in the observations~\cite{MATLAB}. The training set $\mathcal{D}=\{(\mathbf{x}_i, y_i)|i=1,\cdots,N\}$ with $N$ observations, constrains the available distribution of functions based on Bayes theorem, and the mean of the posterior distribution is used for prediction. The functional forms of $K(\mathbf{x}_i, \mathbf{x}_j)$ and $\mathbf{h}(\mathbf{x})$ can be selected according to the cross-validation performance of the models~\cite{SI}.


\begin{figure}[h!!!]
    \centering
    \includegraphics[width=1\linewidth]{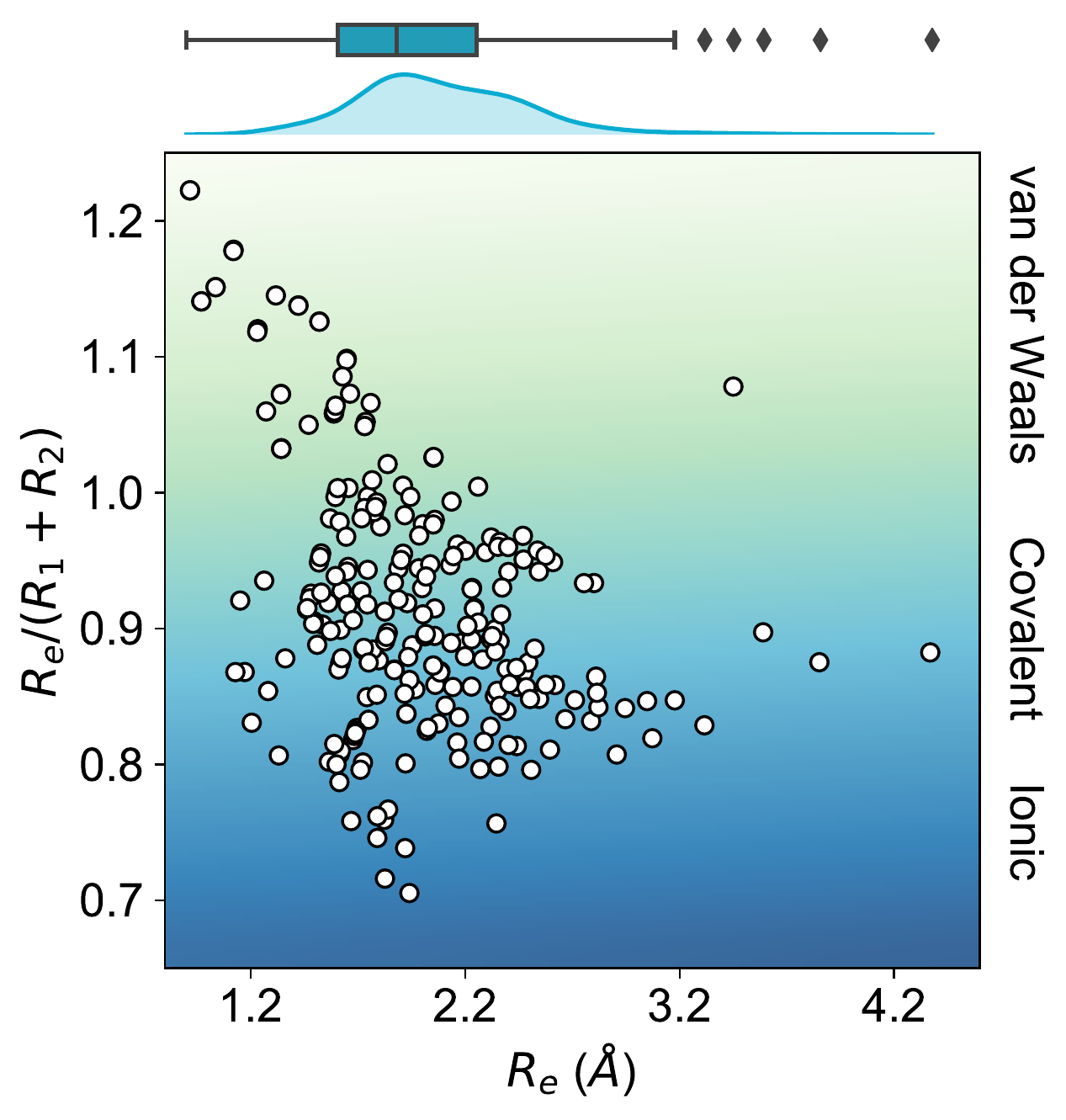}  
    \caption{ Ratio of the equilibrium distance, $R_e$, to the sum of the atomic radii of the atoms forming a molecule, $R_1 + R_2$, vs. $R_e$. The background color indicates the nature of the molecular bond in each of the molecules. The density in the upper part of the figure shows the kernel density distribution of $R_e$. The box plot shows the minimum, the maximum, the sample median, and the first and third quarterlies of $R_e$. The empirical atomic radii of the atoms are taken from Ref.~\cite{slater1964atomic}.}
    \label{Fig:Re_X}
\end{figure}

The employed dataset contains the main spectroscopic constants: $R_e$, $\omega_e$, and $D_0$ for the ground electronic state of heteronuclear diatomic molecules. In particular, it contains the experimental values of $R_e$, $\omega_e$ for 256 heteronuclear diatomic molecules taken from Refs.~\cite{ourweb,Herzberg,Smirnov2008}, whereas the experimentally determined values of $D_0$ are only available for 197 of them. Fig.~\ref{Fig:Re_X} shows the distribution of equilibrium distance of the molecules in the dataset and its ratio to the sum of the atomic radii of the constituent atoms, $R_1+R_2$. Most of the molecules show an equilibrium distance between 1.4~\AA~and 3.8~\AA, with a most probable value of 1.7~\AA. Looking at the values of $R_e/(R_1+R_2)$, it is clear that the molecules within the dataset have different bonds: covalent, van der Waals and ionic. In particular, the present dataset contains a majority of covalent and ionic molecules and only a handful of van der Waals molecules, as shown in Fig.~\ref{Fig:Re_X}.

\begin{figure*}[ht]
    \centering
    \includegraphics[width=1\linewidth]{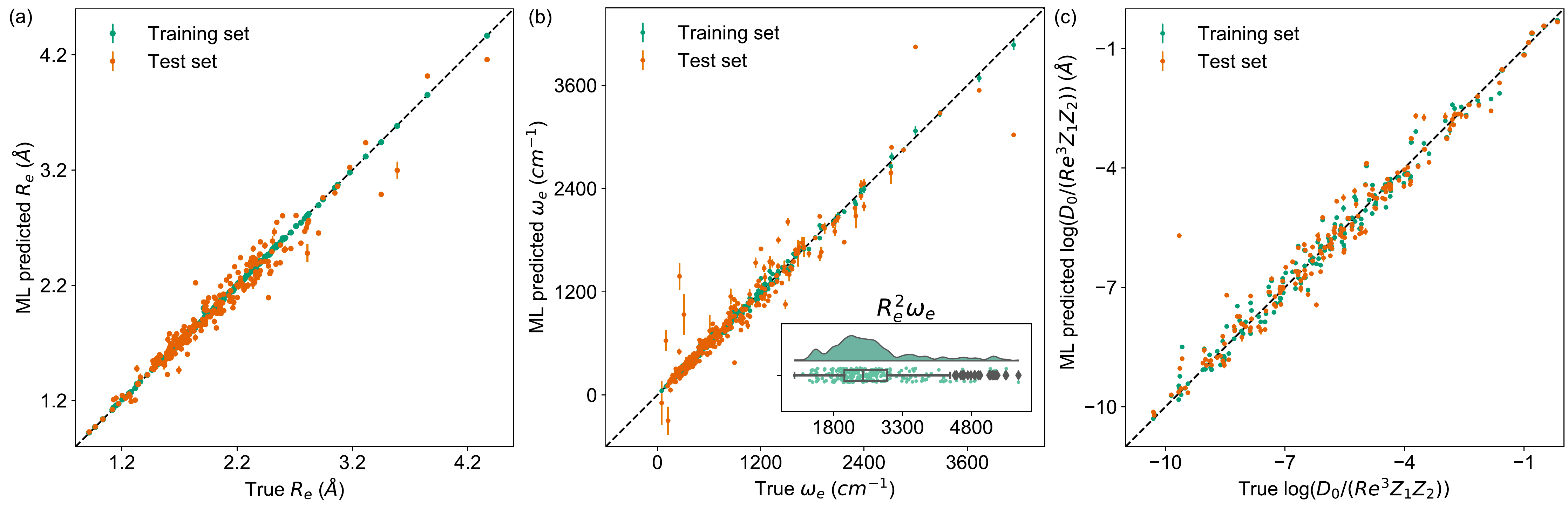}  
    \caption{Predictions of different combinations of spectroscopic constants of diatomic molecules. The values shown are the average of predictions from 1000 MC sampled training/test splittings~\cite{SI}. The GP regression model as learned from the training set gives predictions of the test and training set. Shown are mean and standard derivation of the predictions for each molecule when they are used as training data (shown in green) and test data (shown in orange). (a) Predictions of $R_e$, using the groups and periods of the two atoms, $(g_1,g_2,p_1,p_2)$, as input features. (b) Predictions of $\omega_e$, using $({R^*_e}^{-1},g_1^{iso},g_2^{iso},p_1,p_2)$ as input features. The groups of the atoms $g^{iso}$ also encode the information of hydrogen isotopes, and ${R^*_e}$ is the predicted equilibrium internuclear distance from $(g_1,g_2,p_1,p_2)$. The inset shows the distribution and a box plot of $R_e^2\omega_e$ for the molecules within the dataset. (c) Predictions of $\log(\frac{D_0}{R_e^3 Z_1 Z_2})$, using the predicted equilibrium internuclear distance ${R^*_e}$ and the averages of groups and periods of the two atoms $({R^*_e},\bar{g},\bar{p})$ as input features.}
    \label{Fig:ML_pred_vs_true}
\end{figure*}


Fueled by the idea of the periodicity of the molecules (see, e.g., Ref.~\cite{Hefferlin} and references in it), we use groups, $g_k$, and periods, $p_k$, of the atoms within a molecule, i.e., $k=1,2$, as input features for a GP regression model to predict different combinations of spectroscopic constants, in particular, $R_e$, $\omega_e$ and $\log\frac{D_0}{Re^3 Z_1 Z_2}$, where $Z_k$ stands for the atomic number of the $k$-th atom of the molecule. The last combination of spectroscopic constants was first proposed by Anderson, Parr, and coworkers, by means of the Born-Oppenheimer approximation and assuming that the electron density of an atom within a molecule decays exponentially at the equilibrium distance~\cite{Anderson1971,Anderson1971bis,Anderson1972,gazquez1979universal}. In the GP regression model, as customary in ML, the dataset is divided into training and test sets. However, the present dataset is rather small from a ML perspective. When the dataset is split into training and test sets, the training set may not be representative. This may lead to a bias in the performance of the test set. To solve this problem, we have employed a Monte Carlo (MC) approach, in which the dataset is stratified into $25$ strata based on the level of the true values of the labels ($R_e$, $\omega_e$, and $\log\frac{D_0}{Re^3 Z_1 Z_2}$ in the present work). In each MC step, the training and test data are randomly selected in each stratum. The stratification helps to minimize the change of the proportions of the dataset compositions upon splitting~\cite{raschka2018model}. Additionally, in training the GP regression models, the training sets are further split to perform stratified 5-fold cross-validation for hyperparameter optimization to avoid overly optimistic estimates of the model performance.



The performance of a GP regression model may be quantified by the root mean square error (RMSE) as $\text{RMSE}= \sqrt{\frac{1}{N}\sum_{i=1}^{N}{(y_i - y^*_i)}^2}$, where $y_i$ are the true labels (experimental values) and $y^*_i$ are the predictions, and the normalized error defined as the ratio of the RMSE to the range of the data $r_E = \text{RMSE}/(y_{max}-y_{min})$. 

The GP regression model predictions of $R_e$ as a function of input features $(g_1,g_2,p_1,p_2)$ in comparison with its true values are displayed in panel (a) of Fig.~\ref{Fig:ML_pred_vs_true}, which shows little dispersion of the predicted values with respect to the true values. To further quantify the GP regression model performance we calculate the average RMSE of the predicted $R_e$ on 1000 randomly selected test sets leading to $0.0968~\pm~0.0070$~\AA (Table~\ref{Table:summary}), and $r_E$~=~$2.80~\pm~0.20\%$. The model performance increases as the number of molecules in the training set $N$ grows. It is not yet converged for $N=231$, suggesting that the GP regression model can be further improved by learning from more data in the training set~\cite{SI}.  

\begin{table*}[t]
\centering
\caption{GP regression model predictions of $R_e$, $\omega_e$, and $D_0$. $g_i$ and $p_i$ are the groups and periods of the $i$-th atom, respectively whereas $g^{iso}_i$ stand for the group encoding the information of isotopes of hydrogen \cite{SI}}.

\begin{threeparttable}
\begin{tabular}{p{2cm}p{4.5cm}p{4.5cm}p{4cm}}
\hline
Property                      & Feature                                         & Test RMSE              & Test $r_E$  ($\%$)      \\
\hline

$R_e$   (\AA)                      & ($g_1$, $g_2$, $p_1$, $p_2$)                   & $0.0968 \pm 0.0070$ & $2.80 \pm 0.20 $ \\
$\omega_e$    (cm$^{-1}$)                 & ($R_e^{-1}$, $g_1$, $g_2$, $p_1$, $p_2$)  & $207.2 \pm 2.6$ & $ 5.07 \pm 0.06$          \\
                              & (${R^*_e}^{-1}$, $g_1$, $g_2$, $p_1$, $p_2$)   &  $227.5 \pm 4.6$   & $ 5.56 \pm 0.11$          \\
                              & ($R_e^{-1}$, $g_1^{iso}$, $g_2^{iso}$, $p_1$, $p_2$)  & $ 142.8 \pm 7.0$   & $ 3.49 \pm 0.17$          \\
                              & (${R^*_e}^{-1}$, $g_1^{iso}$, $g_2^{iso}$, $p_1$, $p_2$) & $  176.0 \pm 13.1$  & $ 4.30 \pm 0.32 $          \\
$\log\frac{D_0}{R_e^3Z_1Z_2}$ & ($R_e$, $\bar{g}$, $\bar{p}$)                  &  $0.357 \pm 0.007$    & $ 3.52 \pm 0.07 $          \\
                              & ($R_e^*$, $\bar{g}$, $\bar{p}$)                      & $ 0.451 \pm 0.007 $      & $4.45 \pm 0.07$   \\
\hline
\end{tabular}
    \begin{tablenotes}
        \footnotesize
        \item[a] $R^*_e$ is the predicted value from ($g_1$, $g_2$, $p_1$, $p_2$).
    \end{tablenotes}
\end{threeparttable}
\label{Table:summary}
\end{table*}

Panel (b) of Fig.~\ref{Fig:ML_pred_vs_true} shows the comparison between the predicted $\omega_e$ and its true value. $({R^*_e}^{-1},g_1,g_2,p_1,p_2)$ are used as input features where $R^*_e$ is the predicted equilibrium distance from $(g_1,g_2,p_1,p_2)$. The accuracy of GP regression model for $\omega_e$ is characterized by RMSE~=~$227.5~\pm~4.6$~cm$^{-1}$ and $r_E = 5.56~\pm~0.11\%$. Thus, the GP regression model accurately predicts $\omega_e$, nevertheless for some molecules the prediction is not as good as expected. The overall performance of the GP regression model (Table~\ref{Table:summary}), is improved by utilizing $({R^*_e}^{-1},g_1^{iso},g_2^{iso},p_1,p_2)$ as input features, where $g_k^{iso}$ encodes the information about the hydrogen isotopes of the $k$-th atom in the molecule. Thus, the different isotopologues in the dataset are adequately addressed by input features as explained in the Supplemental Material~\cite{SI}. Further improvement is possible by using the true value of the equilibrium distance, leading to a more precise prediction of $\omega_e$, as displayed in Table~\ref{Table:summary}. Despite the improvement on the description of $\omega_e$ the outliers shown in panel (b) of Fig.~\ref{Fig:ML_pred_vs_true} remain, as the ones in panel (c). Indeed, we have identified the majority of these outliers as bi-alkali molecules, for a detailed description see Supplemental Material~\cite{SI}.

The GP regression model prediction of $\log\frac{D_0}{R_e^3Z_1Z_2}$ vs. its true value is shown in panel (c) of Fig.~\ref{Fig:ML_pred_vs_true}, which shows an outstanding performance. The performance is further supported by an RMSE~=~$ 0.451~\pm~0.007 $ and an $r_E$ equal to $ 4.45~\pm~0.06\%$, as shown in Table~\ref{Table:summary}. In this case, the GP is fed with $({R^*_e},\bar{g},\bar{p})$ as input features, where $\bar{g}=\frac{g_1+g_2}{2}$ and $\bar{p}=\frac{p_1+p_2}{2}$. It is worth mentioning that the accuracy of the GP regression model regression is a factor of 2 better than using standard regression techniques, as shown in the Supplemental Material~\cite{SI}. The learning curve is converged for $n=125$, which suggests that accurate predictions can be made by the models learned from a minimal training set. As a result, with only two atomic properties as input features it is possible to predict the value of $\log\frac{D_0}{R_e^3Z_1Z_2}$ with an accuracy better than $5\%$. The use of the true value for the equilibrium distance reduced the RMSE, as expected, by $20\%$, as shown in Table~\ref{Table:summary}.

From the GP regression models for the three combinations of spectroscopic constants shown in Fig.~\ref{Fig:ML_pred_vs_true}, $\omega_e$ shows the largest error bars regarding the MC sampling technique. This behavior is due to the large variation of the values of $\omega_e$ in every iteration. Indeed, by looking at the distribution of $R_e^2\omega_e$ shown in the inset of panel (b) and the distribution of $R_e$ in Fig.~\ref{Fig:Re_X}, it is clear that $\omega_e$ shows a broad distribution with a multi-modal character. Surprisingly enough, $R_e^2\omega_e=\text{constant}$ seems to be a trend for most of the diatomic molecules, and extensible to any excited state~\cite{Mecke1925,Borkman1968}. However, we do not observe this, as the box plot underneath the inset in panel (b) emphasizes. Indeed, the variation of $R_e^2\omega_e$ may be related to different underlying 2-body potentials for the diatomic molecules considered. Hence, it is another way to show the different bond mechanisms of the molecules within the dataset.  

\begin{figure}[h]
    \centering
    \includegraphics[width=1\linewidth]{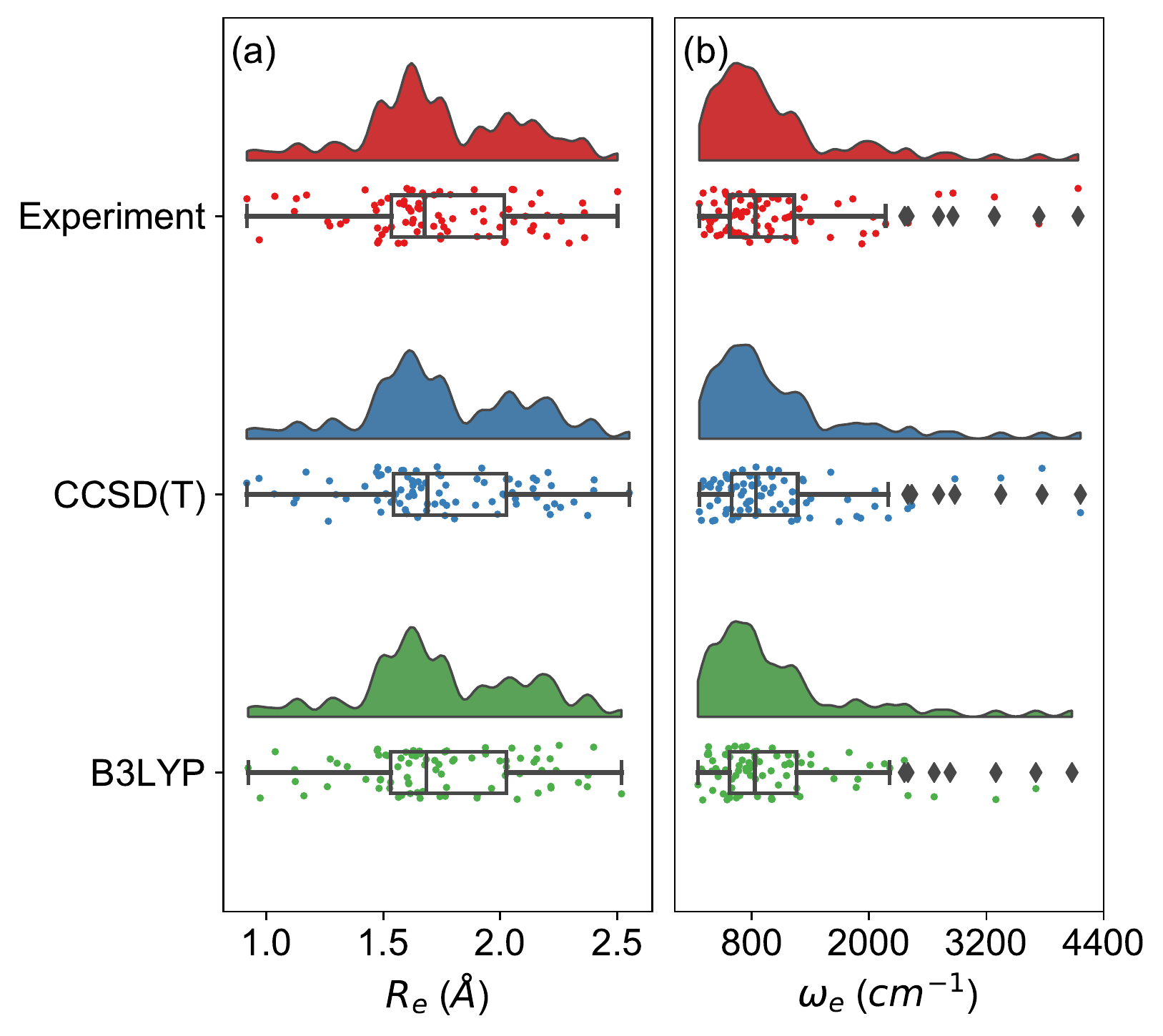}  
    \caption{Comparison of density distributions of experimental and calculated equilibrium internuclear distances $R_e$ and harmonic frequencies $\omega_e$, showing the raw data, probability density, as well as the median, mean, and relevant confidence intervals. }
    \label{Fig:methods_distributions}
\end{figure}

As shown above, the GP regression model for the spectroscopic constants shows a clear universal trend between them. However, the dataset only contains experimentally determined values for spectroscopic constants. Thus, the question arises: do the computationally derived spectroscopic constants lead to the same universal relationships? To answer this question, we have conducted high-level electronic structure calculations using the Molpro package~\cite{molpro,werner2012molpro} with the aug-cc-pV5Z basis set~\cite{pritchard2019new} to calculate the ground state electronic potential energy curve of $81$ molecules in the dataset. The calculations are performed via coupled-cluster with single, double, and perturbative triple excitations CCSD(T)~\cite{ccsdt1989, ccsdt1990} and with the B3LYP~\cite{b3lyp} functional as DFT method. From the potential energy curves, we estimate $R_e$, and $\omega_e$ whose distributions in comparison with the experimental values in the data set are shown in Fig.~\ref{Fig:methods_distributions}. As a result, the distributions of $R_e$ and $\omega_e$ from B3LYP and CCSD(T) show the same features as the experimental one. Therefore, the universal relations for the present dataset are equally fulfilled for high-level electronic structure methods, as it is further elaborated in Supplemental Material~\cite{SI}.

In summary, we have shown, using the GP regression model, that the main spectroscopic constants of diatomic molecules are universally related. This result confirms the scenario that Kratzer and Mecke envisioned a century ago~\cite{Mecke1925,Kratzer1920}. The relationships are independent of the nature of the chemical bond of the diatomic molecule. In particular, we have demonstrated that merely using the group and period of the atoms within a molecule as input features it is possible to predict particular combinations of spectroscopic constants with an error $r_E < 5\%$. In other words, the spectroscopic constants of diatomic molecules can be efficiently learned from an appropriate dataset by a GP regression model, and their values accurately predicted without carrying out quantum chemistry calculations. Besides, the high-level electronic structure calculations ({\it ab initio} and DFT) for the spectroscopic constants show the same distribution as the experimental ones. We conclude that the computationally-derived spectroscopic constants follow the same universal trends as the experimental ones, which are employed in our GP regression model. From our perspective, having reasonable estimates of the main spectroscopic constants will help to optimize the experimental efforts in performing spectroscopy of diatomic molecules. In the same vein, the present work may motivate data science-driven studies on the field of spectroscopy of diatomic molecules. In particular, it will help to evolve the field of spectroscopy towards the current information era.

Finally, we would like to emphasize that there are around $7000$ heteronuclear molecules, and we only utilize $256$ of these for our GP regression model. The limited availability of spectroscopic data (only around $3\%$ of possible heteronuclear diatomic molecules) shows the vast amount of spectroscopy that can be done within the realm of diatomic molecules. The more data we have, the more accurate will be the GP regression model predictions before reaching convergence of the learning curve, and more knowledgeable the community will be about the fundamental properties of diatomic molecules.

We thank Dr. Matthias Rupp for his comments and suggestions and Drs. Daniel Thomas and Uwe Hergenhahn for carefully reading the manuscript.

\bibliographystyle{apsrev}
\bibliography{main.bib}


\end{document}